\documentclass[conference]{IEEEtran}

\IEEEoverridecommandlockouts

\usepackage{cite}
\usepackage{amsmath,amssymb,amsfonts}
\usepackage{algorithm2e}
\usepackage{graphicx}
\usepackage{textcomp}
\usepackage{xcolor}
\usepackage{enumitem}
\usepackage{tabularx}
\usepackage{booktabs}
\usepackage{multirow}
\usepackage{svg}
\usepackage{multicol}
\usepackage{adjustbox}
\usepackage[a4paper, total={184mm,239mm}]{geometry}

\def\BibTeX{{\rm B\kern-.05em{\sc i\kern-.025em b}\kern-.08em
    T\kern-.1667em\lower.7ex\hbox{E}\kern-.125emX}}

\begin{document}
    %\bstctlcite{IEEEexample:BSTcontrol}
    \title{TitanCFI: Toward Enforcing Control-Flow Integrity in the Root-of-Trust \\
%    \thanks{This work was supported by Technology Innovation Institute, Secure Systems Research Center, Abu Dhabi, UAE, PO Box: 9639, and by the Spoke 1 on Future HPC of the Italian Research Center on High-Performance Computing, Big Data and Quantum Computing (ICSC) funded by MUR Mission 4 - Next Generation EU, and by KDT TRISTAN project (g.a. 101095947).}
}

\author{
    \IEEEauthorblockN{
        Emanuele Parisi, Alberto Musa, Simone Manoni, Maicol Ciani, Davide Rossi,\\ Francesco Barchi, Andrea Bartolini, Andrea Acquaviva
    }
    \IEEEauthorblockA{
        \textit{Department of Electrical, Electronic, and Information Engineering (DEI) -- University of Bologna, Italy} \\
        emanuele.parisi@unibo.it
    }
}

\maketitle

\begin{abstract}
Modern RISC-V platforms control and monitor security-critical systems such as industrial controllers and autonomous vehicles.
While these platforms feature a Root-of-Trust (RoT) to store authentication secrets and enable secure boot technologies, they often lack Control-Flow Integrity (CFI) enforcement and are vulnerable to cyber-attacks which divert the control flow of an application to trigger malicious behaviours.
Recent techniques to enforce CFI in RISC-V systems include ISA modifications or custom hardware IPs, all requiring ad-hoc binary toolchains or design of CFI primitives in hardware.
This paper proposes TitanCFI, a novel approach to enforce CFI in the RoT.
TitanCFI modifies the commit stage of the protected core to stream control flow instructions to the RoT and it integrates the CFI enforcement policy in the RoT firmware.
Our approach enables maximum reuse of the hardware resource present in the System-on-Chip (SoC), and it avoids the design of custom IPs and the modification of the compilation toolchain, while exploiting the RoT tamper-proof storage and cryptographic accelerators to secure CFI metadata.
We implemented the proposed architecture on a modern RISC-V SoC along with a return address protection policy in the RoT, and benchmarked area and runtime overhead.
Experimental results show that TitanCFI achieves overhead comparable to SoA hardware CFI solutions for most benchmarks, with lower area overhead, resulting in 1\% of additional area occupation.
\end{abstract}


    \section{Introduction}
\label{sec:introduction}

Open-hardware platforms are becoming widespread in various application domains, including safety and security-critical systems, such as industrial controllers and autonomous vehicles. 
As new System-on-Chip (SoC) designs emerge, a considerable amount of research effort is dedicated to introducing security features to these systems in order to compete with and improve upon mature industrial products.
For this reason, modern systems are equipped with devices such as Trusted Platform Modules (TPM) and Root of Trust (RoT) to support secure boot and firmware signature verification, preventing the execution of malicious code~\cite{ciani2023cyber}. 

While secure boot and firmware verification schemes ensure software authenticity at boot time, embedded systems are often programmed using memory-unsafe languages, which can lead to security vulnerabilities. 
A determined attacker could exploit any read or write vulnerability in the code to bypass traditional memory protection mechanisms~\cite{snow2013jit}\cite{cowan1998stackguard} and alter the control flow of the victim program using Return-Oriented Programming (ROP), or other Code-Reuse Attack (CRA) techniques, aiming to trigger malicious behaviours~\cite{roemer2012return}.

To address the security threats posed by CRAs, modern SoC designs incorporate Control-Flow Integrity (CFI) enforcement policies. 
These policies ensure that an application's control flow adheres to the constraints established by the developer during development and alert the platform runtime if any control-flow violations occur~\cite{abadi2009cfi}.
Various solutions have been proposed to enforce CFI, extending the SoC architecture to provide runtime-checking capabilities for the executed software.
These solutions can be categorized into three main approaches: (i) Instranction Set Architecture (ISA) extensions, (ii) hardware monitors, and (iii) programmable co-processors.
ISA extension techniques~\cite{de2019fixer}\cite{zisslpcfi2023docs} introduce custom opcodes for conducting security checks during program execution. 
While these techniques typically have minimal runtime overhead, they require ad-hoc binary toolchains and invasive modifications to the processor pipeline to accommodate these new instructions. 
Additionally, all software must be rebuilt, and legacy binaries cannot benefit from this protection.
Hardware monitors utilize custom-designed IPs closely integrated with the processor pipeline to monitor control-flow instructions and implement security features like shadow stacks or jump tables directly in silicon~\cite{spang2022dexie}. 
These techniques are transparent to the executed software and do not require extensive modifications to the core architecture. 
However, they lack flexibility in dynamically updating the CFI enforcement policy, requiring the creation of a new monitoring system from scratch.
Security co-processors~\cite{delshadtehrani2022phmon}\cite{delshadtehrani2017nile} rely on a dedicated (semi) programmable core that enables the implementation of the desired policy in software. 
These co-processors use a reserved side-channel to communicate control-flow information. 
This approach allows for the implementation of customized policies on the co-processor, which operates in parallel with the main core. 
However, a co-processor capable of implementing arbitrary policies is typically larger than a customized hardware IP, therefore, there is a significant increase in area overhead.
This paper introduces an innovative co-processor-based architecture for implementing custom CFI policies on a modern RISC-V platform~\cite{ciani2023cyber} that includes the OpenTitan RoT and a RV64GC CVA6 host processor \cite{zaruba2019cost}.
Our approach relies on exploiting the OpenTitan RoT, that is already present on the platform to enable Secure Boot and Remote Attestation, as a CFI co-processor. 
The motivation behind this choice is to harness the RV32IMAC Ibex~\cite{schiavone2017slow} core within OpenTitan to execute custom CFI policies in software. 
This approach avoids the area overhead associated with integrating a separate security monitor and maximizes the utilization of the RoT, which typically remains unused after the platform is initially set up.
Moreover, our solution takes advantage of the security features provided by the RoT, including access to private tamper-proof storage and cryptography accelerators, to provide additional security guarantees with respect to the other state-of-the-art (SoA) solutions.
We demonstrate that our solution incurs negligible area overhead and imposes no, or $<10\%$ overhead for the majority of the tested benchmarks.
Our paper has the following contributions:
\textbf{Design}: We extend the architecture of a modern RISC-V SoC for autonomous vehicle \cite{ciani2023cyber} to (i) enable OpenTitan to observe the stream of control flow instructions retired by the host processor, and (ii) enhance the host core commit stage to filter the control-flow instructions, and enqueue them in a FIFO until the monitor marks them as `safe`.
Moreover, (iii) we extend the firmware running on the OpenTitan IBEX core to analyze the selected instructions and signal any control flow violations.
\textbf{Exploration}: We characterize the overhead of running CFI checks in the RoT by measuring the runtime overhead of our control flow enforcement scheme on a set of benchmarks applications, for different CFI check latencies.
\textbf{Implementation}: We prove the functionality of the designed system implementing a CFI return address protection policy based on a shadow stack. Moreover, we analyze the runtime penalty and hardware overhead of our solution and compare it to the original design.

The paper is organized as follows: Section \ref{sec:related_works} summarizes relevant CFI architectures available in the SoA.
Section \ref{sec:soc_architecture} and \ref{sec:cfi_extensions} introduce the hardware platform and describe the design of TitanCFI.
Section \ref{sec:experimental_results} shows the overhead of TitanCFI compared with the SoA.
Sections \ref{sec:security_implications} and \ref{sec:conclusion} discuss the security advantages of enforcing CFI in the RoT and conclude the paper.
    \section{Related Works}
\label{sec:related_works}

Among the several approaches proposed in the State-of-the-art (SoA) to enforce CFI, the most relevant to our work are (i) FIXER \cite{de2019fixer}, (ii) DExIE \cite{spang2022dexie} and (iii) PHMon \cite{delshadtehrani2022phmon}.

FIXER \cite{de2019fixer} is a security module that enforces CFI by implementing a shadow stack and a jump table \cite{declercq2017survey}.
The host core interfaces FIXER through a set of custom opcodes which signals function calls and returns, and indirect jumps.
FIXER is an example of ISA-based CFI enforcement.
While it has a lightweight execution overhead, FIXER requires access to a custom toolchain and rebuilding all the sources that need to be protected. In contrast, TitanCFI works with a standard toolchain and can protect legacy binaries without the need for rebuilding.

DExIE \cite{spang2022dexie} enforces CFI on the protected host core by implementing a hardware Enforcement Finite State Machines (EFSM) \cite{declercq2017survey}, coupled with a shadow stack.
It represents an example of hardware monitor CFI enforcement.
Similarly to FIXER \cite{de2019fixer} it uses custom-designed hardware IP to enforce the desired CFI policy, to minimize area occupancy and runtime overhead.
Like our approach, it does not rely on any special opcode, and control-flow events are inferred by observing the stream of instructions executed by the host core.
However, the policy requires custom hardware IP and interface toward the main core, whereas our solution uses software-defined policies which use standard bus interconnects.

PHMon \cite{delshadtehrani2022phmon} is a programmable hardware module which belongs to the category of the programmable co-processors.
It consists of a match unit that transparently detects control-flow events by snooping the instructions retired by the host core and a programmable actions unit which enables the user to assign custom actions, like accessing memory or comparing values, to each detected pattern.
Unlike the solution we propose, the action unit is not a fully programmable core, which limits the available CFI policies, and it exploits a custom interconnect to enable the communication between the host core and the hardware monitor.
A significant difference, however, lies in the security guarantees offered by our solution. 
PHMon protects CFI metadata by storing them in a set of virtual memory pages, which the OS marks as reserved, but it lacks a mechanism to ensure the authenticity of such data in the event of an OS breach.
On the other hand, our solution either keeps private data in the tamper-proof storage of the RoT, or, in case of overflow, it exploits the available cryptographic accelerators to ensure authenticity of CFI metadata.
    \section{SoC Architecture}
\label{sec:soc_architecture}

TitanCFI enhances the architecture of a SoA RISC-V platform for micro aerial vehicles \cite{ciani2023cyber}.
The SoC features an host domain based on the CVA6 core, the OpenTitan RoT, and a programmable multi-core accelerator. The programmable accelerator is not considered for the purpose of CFI enforcement.

\subsection{Host Domain Architecture}

The host domain is built around CVA6, a RV64GC, open-source, Linux-capable, RISC-V core \cite{zaruba2019cost}.
The core features a in-order, single-issue, six stage pipeline.
The first two pipeline stages implement the core front-end, which is responsible for (i) generating the next program counter, and (ii) interfacing the instruction cache to fetch the next instruction.
The decode stage (iii) realigns and decodes the instructions and stores them in an issue queue.
The issue stage (iv) contains the issue queue, scoreboard and the Reorder Buffer. It issues instructions to the execute stage once all operands are ready. 
The execute stage (v) contains the functional units, while the commit stage (vi) commits instructions, updates the register file and resolves write-back conflicts through the Reorder Buffer. 
In addition to CVA6, the host domain contains a scratchpad memory, the peripheral subsystem, and a crossbar interconnect which implements the high-bandwidth, low-latency AXI4 protocol. 

\subsection{OpenTitan Root-of-Trust Architecture}

OpenTitan is an open-source silicon RoT that implements a secure enclave aiming at securing sensitive data against hardware attacks, tampering, counterfeiting, and enabling the implementation of security mechanisms such as secure boot and remote attestation.
The hardware architecture of OpenTitan includes (i) a secure microcontroller, (ii) on-board private memory, and (iii) a set of cryptographic accelerators.
The secure microcontroller is Ibex, a open-source RV32IMC MCU optimized for low-gate count, and designed for embedded control applications \cite{schiavone2017slow}.
The cryptographic hardware accelerators efficiently execute compute-intensive security primitives, such as key generation, encryption/decryption, signature generation/verification, and hash calculation. 
OpenTitan is equipped with a 128KB scratchpad SRAM memory and an embedded flash memory enhanced with Error Correcting Code (ECC) and data$\And$address scrambling, for enhanced security and reliability.
In the reference system, OpenTitan is integrated in the SoC and it can access memory through a custom TileLink-to-AXI bridge, as detailed in \cite{ciani2023cyber}. 
Communications between the host domain and the RoT are mediated by a SCMI compliant mailbox.
The mailbox consists of a set of general-purpose memory mapped registers mean for data sharing.
Additionally, it features two registers, named Doorbell and Completion, which are meant to send an interrupt to the Ibex security microcontroller and to the CVA6 host core.
    \section{CFI Extensions and OpenTitan Firmware}
\label{sec:cfi_extensions}

\begin{figure*}[t]
    \centering
    \includegraphics[width=5.7in]{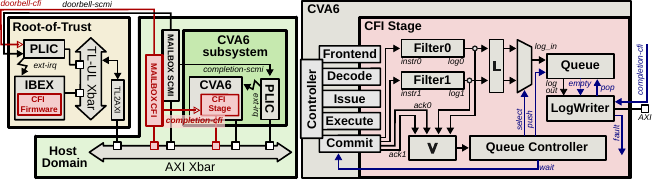}
    \caption{Architecture of TitanCFI. The diagram highlights the proposed architectural modification to the SoC (left) and CVA6 core (right)}
    \label{fig:soc}
    \vspace{-0.2in}
\end{figure*}

This work proposes an enhancement to the reference SoC \cite{ciani2023cyber} and the host CVA6 core \cite{zaruba2019cost} to enable the CFI enforcement in the OpenTitan RoT.
Figure \ref{fig:soc} depicts the major architectural changes.
First, we extend the CVA6 commit stage to (i) filter control-flow operations out of the stream of retired instructions, (ii) extract relevant metadata from the host core scoreboard, (iii) enqueue and forward them towards the RoT. 
Then, we expand the SoC communication system by adding a shared memory region dedicated to exchanging data between CVA6 and the RoT.
Finally, we develop a firmware to enforce CFI.

\subsection{SoC Modifications --- CFI Mailbox}

CFI metadata extracted from the retired instructions are stored in a CFI Mailbox, where they can be read from the RoT.
The design of the CFI Mailbox is analogous to the SCMI-like mailbox already present in the reference SoC \cite{ciani2023cyber}.
We parametrize the general-purpose registers to be large enough to store the CFI metadata required to represent a single control flow instruction.
When a new metadata is ready to be read, the enhanced CVA6 commit stage sets the doorbell register in order to trigger an interrupt in the RoT.
Different from a regular SCMI-like mailbox, the completion register is not connected to the host domain interrupt controller, but directly to the commit stage of the CVA6 core. To indicate that a previously retired instruction has been checked, the result of the CFI enforcement policy can be read from the mailbox, signaling that the RoT is ready to read the next commit log.

\subsection{Host Core Modifications}

\subsubsection{CFI Filters}

The commit stage of the CVA6 core is extended to scan all the instructions retired by each commit port, and to select the control flow operations that need to be checked.
Such operations are indirect jumps, function returns, and function calls.
TitanCFI instanciates two CFI filters, one for each commit port.
A CFI Filter takes a scoreboard entry as input, which is emitted by the commit port, and generates a commit log.
A valid scoreboard entry represents an issued instruction which has been executed, and it is ready to be retired.
From a scoreboard entry the CFI Filter verifies if the retired instruction is relevant to CFI, and it extracts useful metadata, called the commit log, which is needed for CFI enforcement.
A commit log is a 224 bits packet containing four information: (i) instruction program counter, (ii) the uncompressed binary encoding, (iii) the next address, and (iv) the target address.

\subsubsection{CFI Queue and Queue Controller}

The CFI Queue is a FIFO which stores the commit logs extracted by the CFI Filters.
The Queue Controller controls the CFI Queue push signal and, occasionally, it inhibits the CVA6 commit stage which eventually results in stalling the core.
The Queue Control inhibits the commit stage if the CFI Queue is full, or if more that one commit port retires a control-flow instruction.
This is required since the CFI Queue is implemented as a single entry FIFO, and it does not allow more than one commit log per cycle to be pushed into its internal storage.
Since committing two control-flow instructions in the same cycle is a rare event, we do not expect it to impact the performance of our design.

\subsubsection{CFI Log Writer}

The CFI Log Writer module implements a Finite State Machine (FSM) which pops commit logs from CFI Queue, and writes them to the CFI Mailbox through the SoC interconnect.
When the FSM is idle, it waits for the CFI Queue to contain at least one commit log, and for the CFI Mailbox to be ready.
Then, the Log Writer retrieves a commit log from the queue, divides it into data chunks of equal size, matching the interconnect data bus, which is 64 bits in our case, and initiates AXI transactions to transmit the commit log to the CFI Mailbox.
The final AXI transaction sets the doorbell interrupt register and transitions the FSM into a waiting state and it remains there until the completion signal is asserted by the RoT firmware.
Once the completion signal is received, the FSM reads the result of the CFI enforcement check from the CFI Mailbox and triggers an exception if any control flow violation is detected. 
Finally, the FSM returns to idle and it is ready to pop a new commit log from the CFI Queue.

\subsection{OpenTitan Firmware Design}
\label{subsec:opentitan_firmware_design}

TitanCFI implements the desired CFI policy in C and embeds it into the interrupt service routine dedicated to the CFI Mailbox.
Every CFI policy obeys the same common structure, which consists of three steps: (i) IRQ entry, (ii) policy enforcement, and (iii) IRQ exit.
When the OpenTitan interrupt controller receives the CFI Mailbox interrupt, it wakes up the Ibex microcontroller, directs it to jump to the CFI interrupt service routine, reads the commit log from the CFI Mailbox, and extracts the commit log that describes the control-flow event to be checked.
Then, it parses the binary encoding of the control flow instruction to understand which control flow event it represents, among function call, function return, or indirect jump, and applies the programmed CFI enforcement policy.
Finally, the Ibex writes a binary value to the first mailbox entry, to signal whether any control-flow violation has occurred, and sets the mailbox IRQ completion register, to signal that the check is over and OpenTitan is ready to read a new commit log from the mailbox.

    \section{Experimental Results}
\label{sec:experimental_results}

\subsection{Experimental environment}

We synthesized TitanCFI on FPGA, and simulated a set of benchmarks to estimate how the architectural modifications we proposed impact hardware utilization and runtime overhead.
FPGA synthesis has been run on Xilinx Vivado 2020.2 targeting the Virtex UltraScale+ VCU118 system.
The benchmarks comes from EmBench-IoT v1.0 and RISC-V-Tests, and were built with a standard RISC-V toolchain featuring GCC 12.2.0 and \texttt{-O3} optimization level.

\subsection{OpenTitan firmware analysis}
\label{subsec:opentitan_firmware_analysis}

\begin{table}[t]
    \centering
    \footnotesize
    \caption{Cycles required to implement the return address protection policy in OpenTitan}
    \label{tab:results_firmware}
    \begin{adjustbox}{width=3.0in,center}
        \begin{tabular}{@{}lllrrrrrrrrr@{}}
            \toprule
                                                  &
            \multirow{2}{*}{Op.}                  & 
                                                  & 
            \multicolumn{3}{c}{Instructions [\#]} & 
            \multicolumn{3}{c}{Cycles [\#]}       & 
            \multicolumn{3}{c}{Cycles [\%]}       \\
    
            \cmidrule(lr){4-6}
            \cmidrule(lr){7-9}
            \cmidrule(lr){10-12}
            & & &
            IRQ & CFI & TOT & 
            IRQ & CFI & TOT & 
            IRQ & CFI & TOT \\
    
            \midrule

            \multirow{8}{*}{\rotatebox[origin=c]{90}{IRQ}}  &
            \multirow{4}{*}{\rotatebox[origin=c]{90}{CALL}} &
                Logic    &  8  & 15 & 23 &  59 &  27 &  86 &  23 &  10 &  33 \\
            & & Mem. RoT & 14  &  5 & 19 &  74 &  28 & 102 &  29 &   9 &  38 \\
            & & Mem. SoC &  2  &  4 &  6 &  22 &  48 &  70 &  10 &  19 &  29 \\
            \cmidrule{3-12}
            & & TOT      & 24  & 24 & 48 & 155 & 103 & 258 &  62 &  38 & 100 \\
            \cmidrule{2-12}
                                                            &
            \multirow{4}{*}{\rotatebox[origin=c]{90}{RET.}} &
                Logic    &  8  & 15 & 33 &  59 &  45 & 104 &  21 &  16 &  37 \\
            & & Mem. RoT & 14  &  5 & 19 &  74 &  28 & 102 &  27 &  10 &  37 \\
            & & Mem. SoC &  2  &  4 &  6 &  22 &  48 &  70 &   9 &  17 &  26 \\
            \cmidrule{3-12}
            & & TOT      & 24  & 34 & 58 & 155 & 121 & 276 &  57 &  43 & 100 \\
            
            \midrule

            \multirow{8}{*}{\rotatebox[origin=c]{90}{Polling}} &
            \multirow{4}{*}{\rotatebox[origin=c]{90}{CALL}}    &
                Logic    & $-$ & 15 & 15 & $-$ & 27  &  27 & $-$ &  26 &  26 \\
            & & Mem. RoT & $-$ &  5 &  5 & $-$ & 28  &  28 & $-$ &  27 &  27 \\
            & & Mem. SoC & $-$ &  4 &  4 & $-$ & 48  &  48 & $-$ &  47 &  47 \\
            \cmidrule{3-12}
            & & TOT      & $-$ & 24 & 24 & $-$ & 103 & 103 & $-$ & 100 & 100 \\
            \cmidrule{2-12}
                                                            &
            \multirow{4}{*}{\rotatebox[origin=c]{90}{RET.}} &
                Logic    & $-$ & 25 & 25 & $-$ & 45  &  45 & $-$ &  37 &  37 \\
            & & Mem. RoT & $-$ &  5 &  5 & $-$ & 28  &  28 & $-$ &  23 &  23 \\
            & & Mem. SoC & $-$ &  4 &  4 & $-$ & 48  &  48 & $-$ &  40 &  40 \\
            \cmidrule{3-12}
            & & TOT      & $-$ & 34 & 34 & $-$ & 121 & 121 & $-$ & 100 & 100 \\

            \midrule

            \multirow{8}{*}{\rotatebox[origin=c]{90}{Optimized}} &
            \multirow{4}{*}{\rotatebox[origin=c]{90}{CALL}}      &
                Logic    & $-$ & 15 & 15 & $-$ & 27  &  27 & $-$ &  42 &  42 \\
            & & Mem. RoT & $-$ &  5 &  5 & $-$ &  5  &   5 & $-$ &  08 &  08 \\
            & & Mem. SoC & $-$ &  4 &  4 & $-$ & 32  &  32 & $-$ &  50 &  50 \\
            \cmidrule{3-12}
            & & TOT      & $-$ & 24 & 24 & $-$ & 64  &  64 & $-$ & 100 & 100 \\
            \cmidrule{2-12}
                                                            &
            \multirow{4}{*}{\rotatebox[origin=c]{90}{RET.}} &
                Logic    & $-$ & 25 & 25 & $-$ & 45  & 45  & $-$ &  55 &  55 \\
            & & Mem. RoT & $-$ &  5 &  5 & $-$ &  5  &  5  & $-$ &  06 &  06 \\
            & & Mem. SoC & $-$ &  4 &  4 & $-$ & 32  & 32  & $-$ &  39 &  39 \\
            \cmidrule{3-11}
            & & TOT      & $-$ & 34 & 34 & $-$ & 82  & 82  & $-$ & 100 & 100 \\
    
            \bottomrule
        \end{tabular}
    \end{adjustbox}
    \vspace{-0.15in}
\end{table}

To measure the overhead of implementing CFI enforcement policies in the RoT we implemented a well-known return address protection policy called shadow-stack \cite{burow2019sok}.
Our shadow-stack implementation parse the instruction binary to distinguish call from return instructions.
In case of a call, the expected return address is extracted from the commit log and pushed into the shadow stack.
If a return is detected, the return address is extracted from the commit log and compared with the value popped from the shadow stack.
Any mismatch is reported as a security violation.
In both scenarios, the shadow stack is checked for overflow or underflow and eventually saved (or restored) from main memory after having being authenticated using the cryptographic accelerators available in OpenTitan.

Table \ref{tab:results_firmware} reports the cost, in terms of instructions and cycles, of implementing the abovementioned policy in OpenTitan.
It separates the instructions into \texttt{IRQ} vs. \texttt{CFI}, and \texttt{Logic} vs. \texttt{Memory-RoT} vs. \texttt{Memory-SoC}.
The former split distinguishes instructions involved in IRQ handling, such as spilling registers to the stack and clearing the interrupt pending bit, from those implementing the CFI policy, such as accessing the mailbox or pushing the return address in the stack.
The latter split considers memory accesses, distinguishing between RoT private scratchpad and SoC memory, from other operations.
The \textit{IRQ} section of Table \ref{tab:results_firmware} shows the cost in cycles of implementing the OpenTitan firmware described in Subsection \ref{subsec:opentitan_firmware_design}.
The results shows that while the cost in cycles of implementing the CFI policy is non negligible, between 258 and 276 cycles per control-flow operation, the large part of it is overhead caused by the OpenTitan micro architecture.
Two major non-idealities emerge from the result analysis: the high ratio of cycles spent in IRQ handling and the overhead imposed by memory accesses.
Approximately 60\% of the number of cycles involved in the implementation of the CFI policy is spent in IRQ handling.
The analysis of the traces produced by the simulator shows that it takes 45 cycles from when the host core set the doorbell interrupt bit in the CFI mailbox to when the Ibex core wakes up from sleep.
Moreover, since accessing the OpenTitan private scratchpad takes approximately 5 cycles per access, and Ibex spill and restore 6 registers, entering and leaving the IRQ results in 105 cycles overhead, independently from the CFI policy implemented.
On the other hand, the number of instruction required to implement the CFI policy logic, between 48 and 58 opcodes, highlight how implementing return address protection in software is feasible and a potentially cheaper alternative to the design of fully customized hardware modules, assuming the RoT architecture is tuned to obtain a sufficiently high IPC.

The sections of Table \ref{tab:results_firmware} labelled \textit{Polling} and \textit{Optimized} report the RoT firmware performance when applying two possible optimizations targetting at optimizing the issues identified in the previous paragraphs.
\textit{Polling} firmware amortizes the cost of IRQ handling by performing some busy waiting cycles polling the doorbell bit of the CFI Mailbox after an instruction has been checked.
With such optimization, the CFI enforcement firmware does not pay the cost of executing expensive IRQ handling procedures and directly executes the CFI enforcement logic, in case a new control flow instruction is already ready in the CFI Queue.
Such optimization requires no hardware modifications, and enforce the desired policy in 112 cycles on average, saving approximately 58\% of overhead with respect to the approach which does not imply busy waiting loops.

A more aggressive optimization, represented by the \textit{Optimized} section of Table \ref{tab:results_firmware}, involves a partial redesign of the OpenTitan internal architecture to reduce the cost of accessing its private scratchpad.
Substituting the internal OpenTitan interconnect with a low-latency interconnect \cite{} would enable accessing the OpenTitan private memory and peripherals in a single cycle and the SoC memory in approximately 8 cycles, instead of 12.
In this last scenario, enforcing return address protection requires 73 cycles on average, more that $70\%$ less than the baseline \textit{IRQ} firmware. 

\subsection{Runtime overhead}

\begin{table}[t]
    \centering
    \footnotesize
    \caption{Runtime slowdown comparison with \cite{spang2022dexie} and \cite{de2019fixer}}
    \label{tab:soa}
    \begin{adjustbox}{width=2.25in,center}
        \begin{tabular}{@{}llrrrrr@{}}
            \toprule

                                              &
            \multirow{2}{*}{Benchmark}        &
            \multicolumn{5}{c}{Slowdown [\%]} \\
            \cmidrule(lr){3-7}
                                  & 
                                  &
            \cite{spang2022dexie} &  
            \cite{de2019fixer}    &
            Opt.                  & 
            Poll.                 & 
            IRQ                   \\
    
            \midrule

            \multirow{4}{*}{\rotatebox[origin=c]{90}{EmBench}}
            & \texttt{aha-mont64}  &   48 & n.a. &  $-$ &  $-$ &  $-$ \\
            & \texttt{edn}         &   47 & n.a. &    1 &    1 &    2 \\
            & \texttt{matmult-int} &   48 & n.a. &  $-$ &  $-$ &    1 \\
            & \texttt{ud}          &   48 & n.a. &   12 &   18 &   43 \\

            \midrule

            \multirow{5}{*}{\rotatebox[origin=c]{90}{RISC-V Tests}}
            & \texttt{rsort}       & n.a. & \multirow{5}{*}{2} &  $-$ &  $-$ &     1 \\
            & \texttt{median}      & n.a. &                    &    3 &    5 &    12 \\
            & \texttt{qsort}       & n.a. &                    &  $-$ &  $-$ &     1 \\
            & \texttt{multiply}    & n.a. &                    &    2 &    3 &     6 \\
            & \texttt{dhrystone}   & n.a. &                    &  360 &  553 &  1318 \\

            \bottomrule
        \end{tabular}
    \end{adjustbox}
    \vspace{-0.15in}
\end{table}

\begin{table}[t]
    \centering
    \footnotesize
    \caption{%
        Statistics and slowdowns of EmBench-IoT and RISC-V Tests.
    }
    \label{tab:benchmarks}
    \begin{adjustbox}{width=3.0in,center}
        \begin{tabular}{@{}llrrrrr@{}}
            \toprule
    
                                             & 
            \multirow{2}{*}{Benchmark}       & 
            \multirow{2}{*}{Cycles}          & 
            \multirow{2}{*}{CF}              & 
            \multicolumn{3}{c}{Slowdown [\%]} \\

            \cmidrule(lr){5-7}
            %& & & & & 73 & 112 & 267 \\
            & & & & Opt. & Poll. & IRQ \\
    
            \midrule

            \multirow{19}{*}{\rotatebox[origin=c]{90}{EmBench}}
            & \texttt{aha-mont64}     & $2.51\text{E+6}$ & $1.50\text{E+1}$ &  $-$  & $-$  & $-$  \\
            & \texttt{crc32}          & $3.49\text{E+6}$ & $1.50\text{E+1}$ &  $-$  & $-$  & $-$  \\
            & \texttt{cubic}          & $1.10\text{E+6}$ & $2.01\text{E+4}$ &    46 &  107 &  390 \\
            & \texttt{edn}            & $4.23\text{E+6}$ & $3.67\text{E+2}$ &  $-$  & $-$  & $-$  \\
            & \texttt{huffbench}      & $3.49\text{E+6}$ & $2.28\text{E+3}$ &     1 &    3 &   11 \\
            & \texttt{matmult-int}    & $4.69\text{E+6}$ & $2.05\text{E+2}$ &  $-$  & $-$  & $-$  \\
            & \texttt{minver}         & $4.75\text{E+5}$ & $4.50\text{E+3}$ &  $-$  &    7 &  153 \\
            & \texttt{nbody}          & $1.21\text{E+5}$ & $4.29\text{E+3}$ &   163 &  301 &  849 \\
            & \texttt{nettle-aes}     & $5.20\text{E+6}$ & $7.95\text{E+2}$ &  $-$  & $-$  & $-$  \\
            & \texttt{nettle-sha256}  & $4.73\text{E+6}$ & $8.57\text{E+3}$ &     1 &    2 &   11 \\
            & \texttt{nsichneu}       & $5.24\text{E+6}$ & $1.70\text{E+1}$ &  $-$  & $-$  & $-$  \\
            & \texttt{picojpeg}       & $4.97\text{E+6}$ & $2.14\text{E+4}$ &     5 &   15 &   58 \\
            & \texttt{qrduino}        & $4.61\text{E+6}$ & $4.35\text{E+3}$ &  $-$  & $-$  & $-$  \\
            & \texttt{sglib-combined} & $3.67\text{E+6}$ & $2.62\text{E+4}$ &     9 &   32 &  142 \\
            & \texttt{slre}           & $3.57\text{E+6}$ & $6.69\text{E+4}$ &    38 &  110 &  401 \\
            & \texttt{st}             & $1.47\text{E+5}$ & $2.31\text{E+2}$ &  $-$  & $-$  &    2 \\
            & \texttt{statemate}      & $3.22\text{E+6}$ & $2.75\text{E+4}$ &  $-$  & $-$  &  129 \\
            & \texttt{ud}             & $1.87\text{E+6}$ & $2.98\text{E+3}$ &  $-$  & $-$  & $-$  \\
            & \texttt{wikisort}       & $4.38\text{E+5}$ & $7.69\text{E+3}$ &    94 &  158 &  418 \\

            \midrule

            \multirow{14}{*}{\rotatebox[origin=c]{90}{RISC-V Tests}} 
            & \texttt{dhrystone}      & $4.57\text{E+5}$ & $2.25\text{E+4}$ &   260 &   452 &  1215 \\
            & \texttt{median}         & $2.53\text{E+4}$ & $1.10\text{E+1}$ &   $-$ &   $-$ &   $-$ \\
            & \texttt{memcpy}         & $1.20\text{E+5}$ & $1.10\text{E+1}$ &   $-$ &   $-$ &   $-$ \\
            & \texttt{mm}             & $1.41\text{E+6}$ & $2.33\text{E+5}$ &  1108 &  1752 &  4311 \\
            & \texttt{mt-matmul}      & $5.76\text{E+4}$ & $2.38\text{E+2}$ &    11 &    22 &    65 \\
            & \texttt{mt-memcpy}      & $4.08\text{E+5}$ & $1.80\text{E+1}$ &   $-$ &   $-$ &   $-$ \\
            & \texttt{mt-vvadd}       & $1.48\text{E+5}$ & $3.30\text{E+1}$ &   $-$ &   $-$ &   $-$ \\
            & \texttt{multiply}       & $3.72\text{E+4}$ & $9.00\text{E+0}$ &   $-$ &   $-$ &   $-$ \\
            & \texttt{pmp}            & $9.01\text{E+5}$ & $5.90\text{E+1}$ &   $-$ &   $-$ &   $-$ \\
            & \texttt{qsort}          & $2.68\text{E+5}$ & $1.10\text{E+1}$ &   $-$ &   $-$ &   $-$ \\
            & \texttt{rsort}          & $3.32\text{E+5}$ & $1.10\text{E+1}$ &   $-$ &   $-$ &   $-$ \\
            & \texttt{spmv}           & $1.67\text{E+5}$ & $1.10\text{E+1}$ &   $-$ &   $-$ &   $-$ \\
            & \texttt{towers}         & $2.01\text{E+4}$ & $9.00\text{E+0}$ &   $-$ &   $-$ &   $-$ \\
            \bottomrule
        \end{tabular}
    \end{adjustbox}
    \vspace{-0.15in}
\end{table}

We run a set of benchmarks to assess the runtime overhead of software CFI enforcement in the RoT, and compare the results with the SoA.
Slowdown is computed by simulating the RTL of the reference SoC \cite{ciani2023cyber} and extracting the cycle-accurate execution trace reporting the number of cycles required to commit each instruction.
Then, we feed the obtained traces to a trace-driven model which emulates the latency required for CFI enforcement.
We emulated three different latencies, matching the results obtained by the firmware analysis reported in Table \ref{tab:results_firmware}, averaging the number of cycles required to process a function call and a function return: (i) 267 cycles, for the \textit{IRQ} firmware, (ii) 112 cycles, for the \textit{Polling} firmware, and (iii) 73 cycles, for the \textit{Optimized} RoT.

Table \ref{tab:soa} compares the overhead of TitanCFI with the one imposed by two SoA architectures: DExIE \cite{spang2022dexie} and FIXER \cite{de2019fixer}.
In both cases, to establish a fair comparison, we constrained the CFI Queue to have depth 1, to emulate the behaviour of stalling the core as soon as a single control flow instruction is retired.
The first section of Table \ref{tab:soa} shows that TitanCFI imposes a runtime overhead lower than the one imposed by DExIE \cite{spang2022dexie}, for which we considered the best runtime overhead obtained fetching data from the relative paper.
Notice that the authors of \cite{spang2022dexie} report a reduction in the clock frequency of the tested cores, when interfaced with the DExIE security monitor.
While compensating for clock reduction makes the overhead of DExIE \cite{spang2022dexie} negligible, Table \ref{tab:soa} highlights how enforcing CFI in the RoT leads to minimal overhead with respect to custom hardware monitor in 3 out of 4 benchmarks tested by \cite{spang2022dexie}.

Comparing TitanCFI with FIXER \cite{de2019fixer} is more complicated because the authors of \cite{de2019fixer} reports a $1.5\%$ runtime overhead for their solution, without showing the breakdown of how such results were obtained.
We observe that, with the exception of the \texttt{dhrystone} benchmark, TitanCFI has a mean overhead of $~5\%$ in the worst case, only marginally higher than the result reported by the authors of \cite{de2019fixer}.

While comparing the overhead of TitanCFI with the results obtained by \cite{spang2022dexie} and \cite{de2019fixer} is useful to check how it compares with alternative SoA architecture for CFI enforcement, it does not provide a clear overview on the real performance degradation we can expect to get in a real-world scenario.
We see two major criticalities: (i) only functions with a moderately low number of control-flow instructions are chosen, and (ii) the benchmark used are normally used to assess core performance, and they are not relevant to the problem of CFI enforcement.
While this work, like others in the SoA, does not face the second issue, it solves the first criticality by computing the overhead of TitanCFI on a larger pool of benchmarks, considering the whole EmBench-IoT suite, and most of RISC-V-Tests.
Table \ref{tab:benchmarks} reports, for each benchmark, the number of cycles required to complete the execution, the number of control flow instructions retired, and the slowdown obtained by TitanCFI in the three version described in Subsection \ref{subsec:opentitan_firmware_analysis}, using a CFI queue of size 8.
While our solution requires tens of cycles to enforce CFI, and will require further optimization to be applied to benchmarks  with a high number of control flow instructions, it gives $<10\%$ overhead for most of the kernel tested.
We see two main approaches to improve the performance of TitanCFI.
First, ad-hoc static source code analysis techniques should be developed to recognise hot-spots in the code and there selectively inline function calls to alleviate the pressure on the CFI enforcement module.
Additionally, TitanCFI should be enhanced to enforcing CFI per thread, to selectively protect only the processes exposed at the boundary of the system, dealing with potentially tainted data and inputs.
All the benchmarks in the Table \ref{tab:benchmarks} which impose the highest overhead are not kernel which requires CFI enforcement, since they aim at performing linear algebra or physics computation, and they are not expected to be protected in a real-world scenario.

\subsection{Hardware utilization overhead}

\begin{table}[t]
    \centering
    \footnotesize
    \caption{%
        Hardware resource utilization with respect to DExIE \cite{spang2022dexie}
    }
    \label{tab:area}
    \begin{adjustbox}{width=2.3in,center}
        \begin{tabular}{@{}llrrrr@{}}
            \toprule
            & & w.o CFI & CFI & $\Delta$ & Overhead \\
            
            \midrule
            \multirow{3}{*}{\rotatebox[origin=c]{90}{Host}} 
                & LUT       & $5.02\text{E+4}$ & $5.14\text{E+4}$ & $1.16\text{E+3}$ & $+2.3$ \% \\
                & Registers & $3.04\text{E+4}$ & $3.22\text{E+4}$ & $1.77\text{E+3}$ & $+5.8$ \% \\
                & BRAM      & $6.60\text{E+1}$ & $6.60\text{E+1}$ & $-$              & $-$       \\
            
            \midrule
            \multirow{3}{*}{\rotatebox[origin=c]{90}{SoC}} 
                & LUT       & $4.41\text{E+5}$ & $4.41\text{E+5}$ & $1.33\text{E+3}$ & $+0.3$ \% \\
                & Registers & $2.57\text{E+5}$ & $2.58\text{E+5}$ & $2.19\text{E+3}$ & $+0.9$ \% \\
                & BRAM      & $2.68\text{E+2}$ & $2.68\text{E+2}$ & $-$              & $-$       \\

            \midrule
            \multirow{3}{*}{\cite{spang2022dexie}}
                & LUT       & $4.66\text{E+3}$ & $8.02\text{E+3}$ & $3.36\text{E+3}$ & $+72.1$ \% \\
                & Registers & $3.09\text{E+3}$ & $5.33\text{E+3}$ & $2.24\text{E+3}$ & $+72.5$ \% \\
                & BRAM      & $1.36\text{E+2}$ & $1.42\text{E+2}$ & $6.00\text{E+0}$ & $+4.4$  \% \\

            \bottomrule
        \end{tabular}
    \end{adjustbox}
    \vspace{-0.15in}
\end{table}

From FPGA synthesis we report how many LUTs, registers, and BRAMs TitanCFI requires to evaluate how the architectural modification we propose impacts on the resource utilization of the SoC and of the host core.
Table \ref{tab:area} shows that our architectural modifications have negligible impact on hardware resource utilization, equal to $<1\%$ on the entire SoC, and $<6\%$ considering only the host core.
Comparing with the best result reported by \cite{spang2022dexie}, our solution uses $60\%$ less LUTs, $2\%$ less registers, and does not need any BRAM slice.
Furthermore, our proposed modifications do not impact on the maximum operating frequency of the SoC.
    \section{Security Assumptions and Implications}
\label{sec:security_implications}

Our work is designed to protect software implemented in a memory unsafe language, such as C, which makes it prone to bugs and memory vulnerability that could be exploited.
We assume the CFI Mailbox cannot be tampered by other entities in the SoC.
This is reasonable since other security IPs, such as RISC-V Physical Memory Protection (PMP), can be programmed to inhibit accesses to one or more memory regions so that issuing loads or stores to any address within the protected range results in a access fault exception.
Additionally, we assume the RoT private memory is intrinsically secure and inaccessible by any party other than the Ibex microcontroller.

Implementing CFI enforcement policies in the RoT enables the utilization of its private memory and hardware accelerators.
For example, TitanCFI uses the internal RoT scratchpad to store sensible information, such as the shadow stack.
Many CFI architectures in the SoA \cite{delshadtehrani2022phmon}\cite{delshadtehrani2017nile} store CFI data in reserved pages of virtual memory protected by the Operating System.
TitanCFI promises to enhance such security guarantees since the RoT private memory can never be accessed by any entity in the SoC.
In a multi-process scenario where tens of different processes need to be protected, it is unlikely that all the relevant CFI metadata can live in the RoT private memory at the same time, and occasionally moving data in main memory is required.
A possible solution involves statically assigning a region of main memory to the RoT, using technologies such as RISC-V PMP.
TitanCFI, instead, takes inspiration by Zipper Stack \cite{li2020zipper} and it exploits the OpenTitan cryptographic accelerators to authenticate CFI metadata before spilling data to unsecure locations in main memory.
    \section{Conclusion}
\label{sec:conclusion}

This work introduces TitanCFI, a study to investigate the possibility of implementing CFI enforcement policies in software in the RoT, enabling the possibility of implementing any policy in software, without designing and integrating custom hardware monitors, or modifying the host core pipeline.
Moreover, it indicates how to take advantage of the properties of RoT such as tamper-proof memory and cryptographic accelerators for enhancing the security guarantees provided by the CFI enforcement scheme.
We prove our solution by extending the architecture of a modern RISC-V SoC for secure systems \cite{ciani2023cyber}. We demonstrate the proposed approach exhibits minimal hardware overhead and does not compromise timing.
Tests on EmBench-IoT and RISC-V-Tests benchmark suites show that the majority of benchmarks impose no, or $<10\%$ runtime overhead. 
The overhead of our system is comparable to the one imposed by alternative enforcement schemes in literature \cite{spang2022dexie}\cite{de2019fixer} for most benchmarks. 
Future works will test the proposed approach on alternative RoT and more capable platforms, featuring multi-core hosts, implementing alternative CFI policies.
    
    \bibliographystyle{ieeetr}
    \bibliography{main}

\end{document}